\documentclass[runningheads]{llncs}
\usepackage{graphicx} 

\usepackage[utf8]{inputenc} 
\usepackage[T1]{fontenc}    
\usepackage{hyperref}       
\usepackage{url}            
\usepackage[anythingbreaks]{breakurl}

\usepackage{breakurl}
\usepackage{hyperref}
\usepackage{booktabs}       
\usepackage{amsfonts}       
\usepackage{nicefrac}       
\usepackage{microtype}      
\usepackage{lipsum}
\usepackage{graphicx}
\usepackage{graphics}
\usepackage{amsmath}
\usepackage{cleveref}
\usepackage[skip=1pt, indent=40pt]{parskip}
\usepackage[table]{xcolor}

\DeclareMathOperator*{\argmax}{arg\,max}

\newcommand{\blinded}[1]{{\sc blinded for review}}

\title{Dice, but don't slice: \\
Optimizing the efficiency of ONEAudit}

\author{Jake Spertus\inst{1}\orcidID{0000-0003-0682-7677} \and
Amanda Glazer\inst{2}\orcidID{0000-0002-3229-7924} \and
Philip B.\ Stark\inst{1}\orcidID{0000-0002-3771-9604}}
\authorrunning{J. Spertus et al.}
\institute{University of California, Berkeley \\
\email{\{pbstark,jakespertus\}@berkeley.edu}\\
\url{https://www.berkeley.edu} \and
The University of Texas at Austin \\
\email{amanda.glazer@austin.utexas.edu}\\
\url{https://www.utexas.edu}
} 

\begin{document}

\maketitle

\begin{abstract}
    ONEAudit provides more efficient risk-limiting audits than other extant methods when the voting system cannot report a cast-vote record linked to each cast card.
    It obviates the need for re-scanning; it is simpler and more efficient than `hybrid' audits; 
    and it is far more efficient than batch-level comparison audits. 
    There may be room to improve the efficiency of ONEAudit further by tuning the statistical tests it uses and by using stratified sampling. 
    We show that tuning the tests by optimizing for the reported batch-level tallies or integrating over a distribution reduces expected workloads by 70-85\% compared to the current ONEAudit implementation across a range of simulated elections.
    The improved tests reduce the expected workload to audit the 2024 Mayoral race in San Francisco, California, by half---from about 200 cards to about 100 cards.
    In contrast, stratified sampling does not help: it increases workloads by about 25\% on average. 
\end{abstract}

\section{Introduction}

\emph{Risk-limiting audits} (RLAs)
\cite{stark23}
are a key ingredient in \emph{evidence-based elections}
\cite{starkWagner12,appelStark20}, which produce affirmative evidence that the reported winners really won.
By manually inspecting a random sample from a trustworthy record of validly cast votes \cite{appelEtal20}, an RLA guarantees that wrong reported outcomes\footnote{%
Contest winners, not exact tallies.
} 
will, with high probability, be caught and corrected
before they become final.
If routinely employed, RLAs could support trustworthy elections and earn voters' trust in the electoral process.

This paper aims to reduce barriers to routine RLAs by lowering the workload of ONEAudit \cite{stark2023ONEAudit}
(described later in this section), currently the most flexible, transparent, and efficient RLA method.\footnote{%
    ONEAudit (overstatement-net-equivalent audit) \cite{stark2023ONEAudit} creates reference values for ballot cards from batch or contest totals when cast-vote records---digital records of the machine interpretations of individual ballot cards---are not available from the voting system. 
    That makes it possible to conduct card-level comparison audits even when the voting system cannot export cast-vote records uniquely linked to individual ballot cards.
}
We show that stratifying the sample using batch information does not improve the efficiency of ONEAudit, even when the sharpest known methods for stratified inference are used \cite{spertusEtal24}.
On the other hand, new methods to optimize the bets for testing-by-betting \cite{waudby-smithRamdas24,waudby2025universal} can substantially reduce ONEAudit sample sizes.

The overall cost of an RLA involves a fixed cost of setting up the audit and a variable cost that depends on the number of ballot cards sampled.\footnote{%
A \textit{ballot} is a set of ballot \textit{cards} cast by a single voter. Ballots are typically not kept together as an identifiable unit after being cast. The sampling unit for an RLA is a single card or a batch of cards, typically grouped by voting machine or precinct. 
} 
How the election jurisdiction organizes ballot cards (the sizes of batches, whether cards are imprinted with identifiers, and similar considerations) affects the marginal cost-per-card.
All else equal, smaller samples are less expensive and make it easier to complete an RLA during the canvass period, 
so reducing sample sizes may induce more jurisdictions to conduct RLAs.

When the reported outcome is wrong, an RLA is intended to lead to a full hand count. 
When the reported outcome is correct, the sample size required to conduct an RLA depends on many factors: 
the number and sizes of contests under audit;
the reported tallies in those contests and the rate of errors in those tallies; 
how physical cards are translated into numeric populations of \textit{assorters} and reported outcomes are translated into formal \textit{assertions} about contest winners and losers;
the availability of cast-vote records (CVRs) `linked' to individual identifiable cards;
the sampling design (including the size of the sampling unit, whether sampling probabilities are uniform across units, whether units are sampled with or without replacement, and whether the sample is stratified);
and the risk-measuring functions used to check assertions. 

Some details are dictated by
the voting technology, legal requirements, and the particular
logistics of handling and organizing ballot cards. 
Some are under the control of the auditor, 
and can be chosen to maximize the \textit{efficiency} of the audit---often defined in terms of the expected number of cards that must be checked by hand (e.g., in Spertus \cite{spertus2023cobra} or Spertus et al. \cite{spertusEtal24})---or to enhance the simplicity and transparency of the audit \cite{stark10d,stark2023ONEAudit}.
For instance, batch-level comparison audits may naturally fit existing hardware, auditing practices, or recount laws;  
card-level comparison audits (CLCAs) are more efficient, but require the voting system to produce CVRs linked to identifiable ballot cards;
and stratification may be useful to accommodate laws mandating independent sampling across jurisdictions, enforce workload fairness constraints, or increase efficiency, but it introduces additional complexity into the analysis \cite{ottoboniEtal18,SpertusStark2022,spertusEtal24}. 

The most efficient RLA strategy is a CLCA using style-based sampling \cite{glazerEtal21,glazer2023stylish} and comparison-optimal betting test supermartingales (TSMs) \cite{spertus2023cobra}. 
But that strategy needs the voting system to produce a linked CVR for each card, which is rare in the US. 
For instance, in California votes are counted by a mixture of precinct count optical scan (PCOS) and central count optical scan (CCOS).
Modern voting systems can produce linked CVRs for cards tabulated with CCOS but not for 
cards tabulated with PCOS---a deliberate limitation intended to preserve voter privacy.

Historically, heterogeneous voting equipment has been accommodated by ignoring CVRs and conducting a \textit{card-polling audit} (CPA) \cite{lindemanEtal12}; by using stratified sampling and conducting a \textit{hybrid audit} \cite{ottoboniEtal18,stark20a,SpertusStark2022}; or by using size-weighted sampling to conduct a batch-level comparison audit, treating CVR-linked cards as batches of size one \cite{stark09b}.\footnote{%
Fuller et al.\ \cite{fullerEtal23} proposed a method that
re-scans PCOS cards as needed for the audit, and
Stark \cite{stark23b} proposed a method that imprints PCOS cards with nonces as they are scanned.
To the best of our knowledge, neither method has been used in an actual audit.
}

ONEAudit \cite{stark2023ONEAudit} is a more recent method of conducting RLAs of elections when linked CVRs are not available for every cast card. 
It compares manual interpretations of cards to reference values constructed in a way that is ``overstatement-net-equivalent'' to the reported results:
the reference values yield the correct winner if and only if the reported results identified the correct winner.
This allows ONEAudit to take advantage of linked CVRs where they are available and batch subtotals elsewhere, without the need for stratification.
Stratification may still be useful to accommodate legal\footnote{%
Some states (e.g., California) require jurisdictions to draw their own audit samples independently, which yields a stratified sample for contests that cross jurisdictional boundaries.
Other states (e.g., Colorado) draw audit samples centrally from the state as a whole, without stratification.
} 
or logistical\footnote{%
For instance, stratification may be helpful to allow auditing to start before all cards have been tabulated by partitioning the population of cards into those tabulated as of a particular date and those tabulated later.
}  
requirements,
and potentially to increase efficiency, a possibility explored below.

The efficiency of ONEAudit depends in part on the heterogeneity of the votes in the batches from which the reference values are derived. 
If the reported votes in a batch are all for Alice, then the ONEAudit reference value for every card in the batch exactly matches how it was tabulated; and
if the reported tabulation is correct, every reference value will match the human reading of the votes from the corresponding card.
In that case, ONEAudit is just as efficient as a CLCA that uses linked CVRs from the voting system.
In contrast, if the reported votes were 50\% for Alice and 50\% for Bob, no single reference value can exactly match the human reading of the votes from the corresponding card, even if the batch-level tally was perfectly accurate.
Then, ONEAudit would be less efficient than a CLCA that used linked CVRs from the voting system (if those CVRs existed!), but it is still more efficient than batch-level comparison audits.

The next section formalizes the notation for RLAs, with a particular focus on ONEAudit.
In \Cref{sec:dicing} we review testing-by-betting and propose new betting strategies for ONEAudit.
We review stratified RLAs in \Cref{sec:slicing}. 
\Cref{sec:simulations} uses simulations to evaluate the workload of ONEAudit with different betting and stratification in a range of hypothetical elections. 
\Cref{sec:case_study} estimates the sample sizes required for ONEAudit for the 2024 San Francisco mayoral race. 
\Cref{sec:discussion} concludes with a discussion of our findings and recommendations. 

\section{RLAs via assorters and assertions}
\label{sec:setup}

Under the SHANGRLA framework \cite{stark20a}, RLAs are reduced to testing claims about the means of
lists of non-negative numbers.
The numbers come from the actual votes on ballot cards. 
In particular, an \textit{assorter} $A(\cdot)$ maps 
a vote to a number in $[0,u]$, where $u$ is a known upper bound that depends on the assorter in question. 
(The mapping may depend on auxiliary information available before the audit starts.)
Let $m_i$ denote the \textit{manual vote record} for the $i$th card: the actual votes on card $i$ as a human would read them, following voter intent rules for the jurisdiction.
The true assorter mean is $\bar{A}^m := \frac{1}{N} \sum_{i=1}^N A(m_i)$; it is unknown unless there is a full hand count.
The number of assorters involved in auditing a contest depends on the social choice function for that contest,
the number of candidates, and the audit strategy.

For instance, the winners of a multi-winner plurality contest with $K$ candidates and $W$ winners can be characterized by $W(K-W)$ assorters, one for each (reported winner, reported loser) pair \cite{stark20a}.
Each assorter takes the value 1 if the card has a vote for the reported winner in the pair, the value 0 if it has a vote for the reported loser in the pair, and the value 1/2 otherwise, so $u=1$. 
If the means of all $W(K-W)$ resulting lists are all greater than 1/2, every reported winner really got more votes than every reported loser.

An \emph{assertion} is a claim that an assorter mean is greater than 1/2.
An assertion can be checked by testing its \emph{complementary null hypothesis} that the mean is less than or equal to 1/2: to reject that hypothesis is to conclude that the assertion is true.
For most social choice functions (including all scoring rules), a reported outcome is correct if and only if every assertion in a given set of assertions is true \cite{stark20a,blomEtal21};
but for IRV contests, a reported outcome is correct if and only if every assertion in one of some number of sets of assertions is true \cite{ekEtal23,Ek_2024}.

Consider a single assertion $\bar{A}^m > 1/2$.
Card-polling audits (CPAs, \cite{lindemanEtal12,stark23}) check the assertion directly by sampling values of $A(m_i)$.
Card-level comparison audits (CLCAs, \cite{stark10d,stark23}) and ONEAudit \cite{stark2023ONEAudit} involve the difference between $A(m_i)$ and a `reference value' $r_i$, specified before the audit starts.
For instance, if a CVR $c_i$ is available for card $i$, we could take $r_i = A(c_i)$.
We assume that the reference values satisfy 
$\bar{r} := \frac{1}{N} \sum_{i=1}^N r_i > 1/2$, or equivalently that the 
\textit{reported assorter margin} $v := 2 \bar{r} - 1 > 0$.
(According to the reference values, the reported outcome is not wrong.)
The \textit{overstatement} of $r_i$, i.e., the amount by which it exceeds $A(m_i)$, is $\omega_i := r_i - A(m_i)$. 
The \textit{overstatement assorter} 
is
\begin{equation}
    x_i := O(m_i) = \frac{1 - \omega_i/u}{2  - v/u}, \label{eqn:overstatements}
\end{equation}
an affine transformation of the overstatement.
Let $\bar{x} := \frac{1}{N} \sum_{i=1}^N x_i$. 
Then $\{\bar{x} \leq 1/2\} \iff \{\bar{A}^m \leq 1/2\}$: the two conditions are equivalent.
The advantage of basing an RLA on $\{x_i\}_{i=1}^N$ is that it tends to have lower variance when each reference value $r_i$ is close to its corresponding assorter value $A(m_i)$,
which reduces the sample size 
needed to confirm the reported outcome when it is correct.
If every reference value is equal to its corresponding assorter value, $x_i = (2-v/u)^{-1}$ for every card $i$.
This tends to make CLCAs far more efficient than CPAs when the statistical test is tuned appropriately \cite{spertus2023cobra}. 


For cards for which the voting system does not provide a linked CVR,
$r_i$ can be set arbitrarily (as long as $\bar{r} 1/2$).
ONEAudit derives $r_i$ for those cards using
other information the voting system can export,
such as batch subtotals. 
In the U.S., current voting systems generally can report a CVR `linked' to each cast card for 
vote-by-mail ballots, but can report only groups of CVRs for batches of ballots cast in person (e.g., cast in a particular precinct or scanned by a particular scanner), with no way to tell which CVR in the group corresponds to which card in the batch.
In that situation, a reference value $r_i$ can be derived from the average of $A(c_i)$ (the average of the assorter applied to the CVRs) for cards in the batch.
That is how $\{r_i\}_{i=1}^N$ were derived for in-person ballots in the pilot ONEAudit RLA in San Francisco, CA, discussed below. 

When the reported assorter total for a batch is correct (i.e., when $\sum_{i \in \mbox{batch}} r_i = \sum_{i \in \mbox{batch}} A(b_i)$), the mean of the overstatement assorter in the batch is $(2-v/u)^{-1}$. 
This follows from \Cref{eqn:overstatements}: the overstatement assorter is an affine transformation of the overstatement error $\omega_i$ and the average overstatement in a batch is 0 when the reported assorter total in that batch is correct. 
Thus, when the voting system behaves correctly, the overstatement error 
has the same mean in every batch, but the heterogeneity may differ across batches.
In plurality contests, a batch is more homogeneous when the original reported assorter mean is near 0 (favoring the loser) or 1 (favoring the winner), or when there are many non-votes. 
Constructing batches to reduce heterogeneity (e.g., by reducing batch size) could improve efficiency if the statistical test can take advantage of the lower variance.

\section{Dicing: efficient testing-by-betting for ONEAudit}
\label{sec:dicing}


\textit{Betting test supermartingales} (TSMs)---real-valued stochastic processes that are nonnegative supermartingales starting at 1 if the null hypothesis is true---are the basis of the most efficient, sequentially valid RLAs known \cite{waudby-smithEtal21,stark23,spertus2023cobra}. 

Let $X_i$ denote a random draw from a population on $[0,1]$ with true mean $\mu$.
Define
$$M_t := \prod_{j=1}^t [1 + \lambda_j (X_j - \eta_j)],$$
where $\eta_j$ is (an upper bound on) the mean of the population just before the $i$th draw if 
$H_0: \mu \leq \eta$.
Then $(M_t)_{t \in \mathbb{N}}$ is a TSM for $H_0$.
For sampling uniformly and independently with replacement $\eta_j = \eta$; 
for sampling uniformly without replacement $\eta_j = (\eta - \sum_{\ell=1}^{j-1} X_\ell) / (N-j-1)$. 
The TSM is the accumulated wealth of a gambler after $t$ rounds of betting that the null is true, 
risking $\lambda_j \eta_j  \in [0,1]$ of their wealth on the $j$th round. 
The bets $\lambda_j$ must be nonnegative
to prevent the gambler from profiting 
when the draw is less than $\eta_j$ (in which case the game would be favorable rather than fair or sub-fair when the null is true), and cannot exceed $1/\eta_j$ or the gambler might go into debt, which is forbidden. 
The $j$th bet can depend on history $(X_1, \ldots, X_{j-1})$ but not on $X_k$ for any $k \ge j$: it must be \textit{predictable} or \emph{non-anticipating}.
Ville's inequality \cite{ville39} implies that $P_t := 1 \wedge (1/M_t) \in [0,1]$ is a sequentially-valid $P$-value for $H_0$;
that is, if the null hypothesis is true, the chance that $\inf_t (1/M_t) \le p$ is at most $p$, for any $p \in [0, 1]$.

An RLA is efficient if it stops after examining relatively few ballots when the reported outcome is correct.
We measure efficiency by the expected 
value and 90th percentile of the number of cards the audit examines before stopping. 
For a given set of votes and a given set of assertions, efficiency depends on how the bets $\lambda_i$ are selected. 
Stark \cite{stark23} and Waudby-Smith et al.\ \cite{waudby-smithEtal21} discuss how to select $\lambda_i$ for CPAs.
The resulting tests are nearly equivalent to the Bernoulli sequential probability ratio test derived by Wald \cite{wald45}, but use a predictable or \textit{a priori} ``plug-in'' estimate of the unknown alternative mean.
Spertus \cite{spertus2023cobra} describes optimal betting for CLCAs, and presents predictable approximations. 
ONEAudit generates far more complicated assorter populations, even for simple assertions (e.g., two-candidate plurality), and efficient betting strategies for ONEAudit have yet to be explored.
That is the topic we address next.

\subsection{Kelly-optimality and \textit{a priori} betting}

Past implementations of ONEAudit \cite{stark2023ONEAudit} set the bets implicitly using a truncated shrinkage estimator of the assorter mean (`shrink-trunc' in ALPHA \cite{stark20a}) or explicitly using COBRA \cite{spertus2023cobra}.
Both seek to approximate \emph{Kelly optimality}, which maximizes the expected log growth of the TSM at time $j$ for a particular population:
$$\lambda_j^* := \argmax_{\lambda \in [0,1/\eta_j]} \mathbb{E} \left \{ \log [1 + \lambda (X_j - \eta_j)] \right \}.
$$
The expected value is with respect to the probability distribution induced by the design, for the assumed population. 
The \textit{oracle} Kelly bet maximizes the expected log growth for the true population of assorter values, which is unknown in practice.
An \textit{a priori} Kelly bet maximizes the log growth for an assumed population derived from the reported tallies \cite{waudby-smithEtal21};
this betting scheme is efficient when those tallies are (approximately) correct.
The truncated shrinkage estimate bet and COBRA are closed-form, approximately Kelly-optimal bets for CPA and CLCA populations, respectively, but bets optimized for those populations are not efficient for ONEAudit overstatement populations, as demonstrated below.

For a generic population and sampling design, the Kelly bet can be found numerically.
In particular, if we \textit{postulate} a finite-population $\{\tilde{x}_i\}_{i=1}^N$ and cards are drawn IID (rather than without replacement), then the \emph{a priori} Kelly bet solves
\begin{equation}
    0 = \frac{d}{d\lambda} \mathbb{E} \log [1 + \lambda (\tilde{X}_i - \eta)]  = \sum_{i=1}^N \frac{\tilde{x}_i - \eta}{1 + \lambda (\tilde{x}_i - \eta)}, \label{eqn:deriv}
\end{equation}
and $\lambda^*$ can be found by bisection search, for instance. 
If the distributions of $\{x_i\}_{i=1}^N$ and $\{\tilde{x}_i\}_{i=1}^N$ are the same, the \emph{a priori} Kelly bet is the oracle Kelly bet.
However, if the distribution of $\{\tilde{x}_i\}_{i=1}^N$ differs substantially from the distribution of the true values $\{x_i\}_{i=1}^N$, the resulting bets may be far from optimal.

The same reported tallies and CVRs used to construct the reference values for the assorters may provide a good initial guess of the population of overstatements. 
One could use that guess to find the \emph{a priori} Kelly bet, and commit to this bet throughout the audit. 
If the reported tallies have large errors (but the reported outcomes are still correct) this could be worse than strategies that do not rely on the reported results, because it anchors the bets to an inefficient choice.

\subsection{Approximately optimal predictable betting: AGRAPA}

Waudby-Smith and Ramdas \cite{waudby-smithRamdas24} note that the strategy in \Cref{eqn:deriv} could be made predictable by updating $\lambda_t$ at each time step, replacing $\tilde{x}_i$ with the sample $X_i$ and $N$ with $(t - 1)$. 
This requires $t$ numerical optimizations, one for each time step.
A more tractable bet comes from a Taylor series approximation to the derivative. 
The resulting strategy, AGRAPA (approximate growth rate adapted to the particular alternative), uses the bets
\begin{equation}
    \lambda^{\mbox {\tiny ag}}_t := 0 \vee \frac{\bar{X}_{t-1} - \eta}{\hat{\sigma}_{t-1}^2 + (\bar{X}_{t-1} - \eta)^2} \wedge \frac{c}{\eta}, \label{eqn:AGRAPA}
\end{equation}
where $\bar{X}_{t-1}$ is the lagged sample mean, $\hat{\sigma}_{t-1}^2$ is the lagged sample variance, and $c \in [0,1]$ is a user-chosen constant. 
The bet
$\lambda^{\mbox {\tiny ag}}_t$ is larger when the sample mean is further above $\eta$ (the payoff is expected to be higher) and when the sample variance is smaller (the bet is closer to a sure thing). 
Overstatement assorter values for cards with linked CVRs have low variance if the CVRs are accurate.
If most of the cards have linked CVRs, it may be efficient to set $c := 1-\epsilon$ for some small $\epsilon$ (e.g., $\epsilon = 0.05$); this lets the bet become aggressive without betting the whole farm \cite{spertus2023cobra}.

\subsection{Strength in diversity: universal portfolios}
\label{sec:universal_portfolio}
The seminal work of Cover \cite{cover91} suggests averaging across a set of bets rather than maximizing over allowed bets. 
The ($f$-weighted) ``universal portfolio'' bet is defined as:
$$\lambda_t^{\mbox{\tiny up}} := \eta^{-1}\left [ \frac{\int_0^1 \gamma \prod_{i=1}^{t-1} [1 + \gamma (X_i/\eta - 1)] f_\beta(\gamma) d\gamma }{\int_0^1 \prod_{i=1}^{t-1} [1 + \gamma (X_i/\eta - 1)] f_\beta(\gamma) d\gamma } \right ].$$
For a mixing density $f$ on $[0,1]$. 
Cover and Ordentlich \cite{CoverOrdentlich1996} study taking $f$ to be a $\mbox{Beta}(1/2,1/2)$ density or the uniform density. 
We take $f$ to be uniform.
Waudby-Smith et al.\ \cite{waudby2025universal} show that for sampling with replacement, this choice asymptotically does almost as well as the oracle Kelly bet. 
Calculating $\lambda_t^{ \mbox{\tiny up}}$ is computationally expensive and numerically unstable, involving the ratio of integrals of high-order polynomials at every time step $t$.
But there is an inexpensive approximation to the resulting TSM under $\lambda_t^{\mbox{\tiny up}}$. 
Let $\gamma := \eta \lambda \in [0,1]$, then:
$$
    \prod_{i=1}^t [1 + \lambda_i^{\mbox{\tiny up}} (X_i - \eta) ] =  \int_0^{1/\eta} \prod_{i=1}^t [1 + \lambda (X_i - \eta) ] d\lambda  \approx \frac{1}{D} \sum_{j=1}^D \prod_{i=1}^t [1 + \lambda_j (X_i - \eta)],
$$
where $\{\lambda_j\}_{j=1}^D$ is a length-$D$ equispaced grid on $[0, 1/\eta]$.
In the form appearing in the middle, this strategy was suggested for nonparametric testing by Kaplan \cite{kaplan87} a few years before Cover's seminal paper; it is also analogous to Wald's mixture SPRT \cite{wald47} and Robbin's tests-of-power-one \cite{robbinsSiegmund74}.
In words, the TSM with the universal portfolio bet is equivalent to a mixture of TSMs over a uniform distribution on bets. 
This is analogous to the gambler splitting up their initial wealth, placing an equal amount on each possible bet, and summing up their total winnings across the bets.
That this technique is nearly as efficient as the Kelly-optimal strategy follows intuitively: the sum of terms growing exponentially is dominated by the terms with the largest growth rates, and the Kelly-optimal strategy, which maximizes that rate, is one of those terms.

We do not compute the integral exactly but approximate it by a Riemann sum with $D$ terms; $D$ controls the accuracy of the approximation and the computational burden.
This is similar to the various discrete mixture strategies proposed in 
Waudby-Smith et al.\ \cite{waudby-smithEtal21}, Spertus \cite{spertus2023cobra}, and 
Waudby-Smith and Ramdas \cite{waudby-smithRamdas24}, and simpler than refining the grid or adaptively re-balancing the portfolio as $t$ grows \cite{waudby-smithRamdas24,Ek_2024}. 
A fixed, relatively small value of $D$ (say, $D=100$) is adequate
for RLAs, where the sample sizes are typically modest. 
In other problems, a rough discretization might have a large impact on performance, for instance, if we cared about maximizing wealth as $t \rightarrow \infty$, in which case the wealth could become quite peaked around some $\lambda$ that is not included in the grid. 
There is extensive research on  approximating the universal portfolio more accurately \cite{kalai2002efficient}, but for our purposes, the juice is not worth the squeeze. 

\section{Slicing: stratification}
\label{sec:slicing}

ONEAudit obviates the need for stratified hybrid audits, which partition the population into two strata based on the voting technology, draw cards independently across strata, and measure the risk using a stratified test 
\cite{ottoboniEtal18,SpertusStark2022,spertusEtal24}.
Union-of-intersections test sequences (UI-TSs) provide the sharpest known method for sequential stratified testing \cite{spertusEtal24}.
They partition the complementary null into a union of intersections, each of which is tested 
using a product across strata of betting TSMs constructed separately within individual strata.
In theory, an optimal UI-TS is the product of Kelly-optimal TSMs within each stratum.
In practice, finding that UI-TS is computationally prohibitive and the oracle Kelly bet is unknown. 
For any simple alternative, the Kelly-optimal UI-TS can be approximated by partitioning the null into convex regions. 
For each region, the Kelly bet is found for a ``representative'' intersection null (e.g., the centroid of the region); that bet is used for all TSMs within the region.
The TSMs are concave over the intersection nulls within each region, so only the vertices of the regions need to be checked in order to compute and approximately Kelly-optimal UI-TS. 

Spertus et al.\ \cite{spertusEtal24} show that a stratified Kelly-optimal UI-TS is more efficient than an unstratified Kelly-optimal betting TSM when relatively low variance strata have relatively different means.
This mirrors the familiar fact that stratification is helpful in reducing mean-squared error when the between-stratum variance is high and the within-stratum variance is low \cite{cochran77}.
We might suspect that stratifying on batch information---such as the reported assorter mean of the batch---could increase the efficiency of ONEAudit, but \Cref{sec:setup} shows that if the CVRs are error-free, every batch has the same mean; only the variance will differ across batches.
We expect the reported tallies to be quite accurate unless something in the election has gone seriously wrong. 
Stratification is most likely to be advantageous for ONEAudit if it can increase the efficiency of a UI-TS over an unstratified TSM when the \textit{variability} differs in different strata even when the stratum means are equal.

\section{Simulations}
\label{sec:simulations}

We simulated RLAs with a 5\% risk limit to evaluate whether stratification or new betting strategies improve the efficiency of ONEAudit. 
The simulations rescale the original population of overstatements $\{x_i\}_{i=1}^N$ and null mean by $2u/(2u - v)$ to map the population and null means to $[0,1]$.
Simulations were conducted in Python 3.13.2; code is available at 
\url{https://github.com/spertus/UI-TS}.

\subsection{Stratification}

Overstatement populations were constructed with plurality overstatement assorters derived from 10,000 cards with CVRs and 10,000 cards spread across 10 reporting batches, each containing 1,000 cards.
Every card had a valid vote. 
(Sample sizes generally increase with more invalid votes (e.g., Stark \cite{stark2023ONEAudit}), but we do not expect the relative performance of methods to change.)

The global reported assorter mean was $\bar{A}^c \in \{0.505, 0.51, 0.55, 0.60\}$. 
The stratum-level means were parameterized by the ``across gap'' $\Delta_{\mbox{\footnotesize a}} \in \{0.0, 0.5\}$, where $\Delta_{\mbox{\footnotesize a}} = 0.5$ implies the mean in the batch stratum was $\bar{A}^c - 0.25$ and the mean in the CVR stratum was $\bar{A}^c + 0.25$. 
The batch-level means in the batch stratum were parameterized by the ``within gap'' $\Delta_{\mbox{\footnotesize w}} \in \{0.0, 0.5\}$, where $\Delta_{\mbox{\footnotesize w}} = 0.5$ implies the batch level means ranged uniformly between $(\bar{A}^c - \Delta_{\mbox{\footnotesize a}}/2 - \Delta_{\mbox{\footnotesize w}}/2)$ and $(\bar{A}^c - \Delta_{\mbox{\footnotesize a}}/2 + \Delta_{\mbox{\footnotesize w}}/2)$.
The reported tallies were exactly accurate so that $\bar{A}_b^m = \bar{A}_b^c$ for every batch $b$.

We compared RLA sample sizes of an unstratified betting TSM and a stratified betting UI-TS.
The unstratified sample was drawn uniformly with replacement;
the stratified sample was drawn by proportional sampling with replacement and interleaved round-robin, alternating between the two strata. 
The betting TSM used the oracle Kelly-optimal bets, computed by solving \Cref{eqn:deriv} by bisection. 
The UI-TS used the banding strategy described in Spertus et al.\ \cite{spertusEtal24}, partitioning the null into $G = 100$ bands, deriving the oracle Kelly-optimal bets for each stratum at the centroid of each band, evaluating the intersection nulls at the vertices of each band, and minimizing over all 200 vertices. 
Since the bets were fixed within each band, the resulting bets were only approximately Kelly-optimal for each intersection null in the union, but we expect the slack was negligible since the bands were fairly narrow.

\Cref{tab:stratification_results} displays the expected sample sizes of the two approaches.
The Kelly-optimal TSM with unstratified sampling dominates the Kelly-optimal UI-TS with stratified sampling.
The geometric mean of expected sample sizes is about 20\% lower without stratification.

\begin{table}[ht]
\centering
\begin{tabular}{rrr|rrr}
  \hline
$\bar{A}^c$ & $\Delta_{\mbox{\footnotesize a}}$ & $\Delta_{\mbox{\footnotesize w}}$ & Unstratified TSM & ~~Stratified UI-TS & ~~Ratio (\%) \\ 
\hline
0.505 & 0.0 & 0.0 & 18,290 & 20,000 & 91\\ 
  0.505 & 0.0 & 0.5 & 17,428 & 20,000 & 87 \\ 
  0.505 & 0.5 & 0.0 & 15,917 & 20,000 & 80 \\ 
  0.505 & 0.5 & 0.5 & 14,268 & 20,000 & 71 \\ 
  \hline
  0.525 & 0.0 & 0.0 & 1,145 & 1,515 & 76 \\ 
  0.525 & 0.0 & 0.5 & 1,064 & 1,386 & 77 \\ 
  0.525 & 0.5 & 0.0 & 837 & 1,131 & 74 \\ 
  0.525 & 0.5 & 0.5 & 757 & 1,130 & 67 \\ 
  \hline
  0.550 & 0.0 & 0.0 & 327 & 407 & 80 \\ 
  0.550 & 0.0 & 0.5 & 293 & 407 & 72 \\ 
  0.550 & 0.5 & 0.0 & 228 & 286 & 80 \\ 
  0.550 & 0.5 & 0.5 & 216 & 273 & 79 \\ 
  \hline
  0.600 & 0.0 & 0.0 & 85 & 88 & 97\\ 
  0.600 & 0.0 & 0.5 & 71 & 92 & 77\\ 
  0.600 & 0.5 & 0.0 & 59 & 70 & 84 \\ 
  0.600 & 0.5 & 0.5 & 57 & 68 & 84\\ 
   \hline
\end{tabular}
\caption{Estimated expected sample sizes
to confirm that the assorter mean is greater than 1/2 at risk limit 5\% for unstratified and stratified ONEAudit using Kelly-optimal betting strategies. 
There are 20,000 cards in all, 10,000 with CVRs and 10,000 in 100 batches of 1,000 cards, for which ONEAudit assorter values were used to generate populations of overstatements.
$\bar{A}^c$ is the global reported assorter mean; $\Delta_{\mbox{\footnotesize a}}$ is the spread between the assorter mean for the cards with CVRs and the assorter mean of the batches of 1,000 cards; $\Delta_{\mbox{\footnotesize w}}$ is the maximum spread of assorter means across batches. 
For example, in the last row the overall assorter mean is 0.6, the mean for the CVRs is 0.85, the mean for the batches is 0.35, and the 10 batch means are spread evenly between 0.1 and 0.6.
Sampling is with replacement and the sample sizes were capped at 20,000 cards, so these estimates are biased slightly down. 
TSM: test supermartingale; UI-TS: union-of-intersections test sequence.
The last column is the ratio of the TSM sample size to the stratified UI-TS sample size, as a percentage.
Stratification increases the expected sample size in all these numerical experiments.}
\label{tab:stratification_results}
\end{table}

\subsection{Betting}

We compared expected sample sizes for ONEAudit RLAs using different betting strategies.
The populations were again parameterized by the contest-level reported assorter mean $\bar{A}^c$, and heterogeneity parameters $\Delta_{\mbox{\footnotesize a}}$ and $\Delta_{\mbox{\footnotesize w}}$ as described above.
Their values were also the same.
We either let $\bar{A}^m = \bar{A}^c$ or introduced errors into the reported batch tallies (which we expect will generally be more erroneous), setting $\bar{A}^m$ to be half-way between $\bar{A}^c$ and $\eta$, thus halving the true margin.
The simulations involve 100,000 cards in batches, each of which contains 1,000 cards, and 100,000 cards with CVRs. 
Cards were drawn by simple random sampling (without replacement) and the TSMs were computed accordingly. 

Following Waudby-Smith et al.\ \cite{waudby-smithEtal21}, bets were computed as if the sample were drawn IID.
The Kelly-optimal bets were found by solving \Cref{eqn:deriv} over $[0,1/\eta]$ using bisection.
The truncated shrinkage bet involves shrinking the running sample mean toward $\bar{A}^c$, with the default value of the tuning parameter $c$ and $d=20$.
The COBRA bet was computed assuming no 1-vote overstatements and 0.1\% rate of 2-vote overstatements; it was thus fixed just below the maximum bet of $\lambda = 1/\eta$ in most cases. 
AGRAPA was computed as in \Cref{eqn:AGRAPA}, with $c=0.99$.
The universal portfolio wealth was computed using the approximation in \Cref{sec:universal_portfolio}, with the TSM mixed over $D = 100$ equispaced points on $[0,1/\eta]$. 

Results appear in \Cref{tab:betting_stopping_times}. 
AP Kelly had the smallest sample sizes, nearly the same as the oracle Kelly bets, even when the reported tallies were wrong. 
Universal portfolio was the next best, followed closely by AGRAPA.
COBRA was usually the worst bet, except when the mean was far enough above the null that the oracle Kelly bet was near $1/\eta$. 
The bet based on the truncated shrinkage estimate also had poor performance: it is typically too conservative since it assumes, incorrectly, that the variance is maximal, as it would be for a Bernoulli population distribution. 
The geometric mean of the workload ratios comparing AP Kelly to other bets suggests AP Kelly could on average produce workloads that are 91\% smaller than the COBRA bet, 74\% smaller than the bet based on the truncated shrinkage estimate, 31\% smaller than AGRAPA, and 25\% smaller than the universal portfolio bet.
Similarly, the universal portfolio was about 88\% smaller than the COBRA bet, 66\% smaller than the truncated shrinkage estimate bet, and 9\% smaller than AGRAPA, on average.

\begin{table}[htbp]
\centering
\scalebox{0.65}{
\begin{tabular}{rrrr|rrrrrr}
  \hline
$\bar{A}^c$ & $\bar{A}^m$ & $\Delta_{\mbox{\footnotesize a}}$ & $\Delta_{\mbox{\footnotesize w}}$ & Oracle Kelly & AP Kelly & Universal Portfolio & AGRAPA & truncated shrinkage & COBRA \\ 
\hline
0.505 & 0.502 & 0.0 & 0.0 & 73,202 (115,922) & 71,493 (113,043) & 103,786 (149,880) & 106,304 (154,946) & 158,008 (189,423) & 191,042 (199,476) \\ 
  0.505 & 0.502 & 0.0 & 0.5 & 67,277 (107,638) & 67,760 (109,147) & 93,319 (143,173) & 98,171 (147,857) & 160,152 (189,269) & 189,330 (199,511) \\ 
  0.505 & 0.502 & 0.5 & 0.0 & 58,016 (96,992) & 59,136 (96,467) & 85,558 (131,639) & 82,390 (131,582) & 157,225 (186,797) & 187,403 (199,044) \\ 
  0.505 & 0.502 & 0.5 & 0.5 & 55,133 (94,712) & 52,792 (87,743) & 75,683 (123,296) & 78,644 (125,641) & 155,773 (185,975) & 189,746 (199,002) \\ 
  \hline
  0.505 & 0.505 & 0.0 & 0.0 & 25,941 (49,347) & 25,211 (48,603) & 38,689 (68,893) & 40,921 (71,757) & 76,194 (115,980) & 185,968 (196,804) \\ 
  0.505 & 0.505 & 0.0 & 0.5 & 24,247 (45,459) & 23,496 (43,965) & 37,553 (67,685) & 38,654 (70,340) & 76,887 (114,366) & 186,269 (199,014) \\ 
  0.505 & 0.505 & 0.5 & 0.0 & 18,697 (34,957) & 20,118 (37,337) & 30,793 (57,850) & 28,742 (53,071) & 74,181 (107,468) & 182,898 (194,580) \\ 
  0.505 & 0.505 & 0.5 & 0.5 & 17,292 (32,493) & 18,461 (34,509) & 28,033 (52,630) & 26,052 (47,518) & 73,300 (104,434) & 181,746 (194,394) \\ 
  \hline
  0.510 & 0.505 & 0.0 & 0.0 & 24,118 (46,033) & 25,225 (46,615) & 39,721 (71,015) & 42,119 (74,312) & 78,111 (117,215) & 186,123 (196,758) \\ 
  0.510 & 0.505 & 0.0 & 0.5 & 23,088 (43,063) & 22,713 (43,211) & 37,572 (67,753) & 37,964 (69,282) & 76,050 (114,675) & 188,060 (199,015) \\ 
  0.510 & 0.505 & 0.5 & 0.0 & 18,743 (34,358) & 18,943 (36,126) & 29,915 (57,553) & 28,649 (54,976) & 76,199 (109,565) & 182,339 (194,401) \\ 
  0.510 & 0.505 & 0.5 & 0.5 & 17,542 (34,520) & 16,851 (31,960) & 27,151 (51,861) & 26,973 (50,630) & 74,457 (106,746) & 183,134 (194,178) \\ 
  \hline
  0.510 & 0.510 & 0.0 & 0.0 & 7,080 (13,844) & 7,156 (14,092) & 11,157 (22,204) & 12,161 (22,733) & 24,448 (38,449) & 177,439 (188,711) \\ 
  0.510 & 0.510 & 0.0 & 0.5 & 6,479 (12,989) & 6,521 (12,677) & 9,969 (19,450) & 11,302 (21,740) & 23,474 (37,470) & 182,562 (198,021) \\ 
  0.510 & 0.510 & 0.5 & 0.5 & 4,889 (9,818) & 4,547 (9,074) & 6,873 (13,881) & 7,232 (14,328) & 23,127 (35,595) & 160,778 (180,462) \\ 
  0.510 & 0.510 & 0.5 & 0.0 & 5,659 (11,478) & 5,638 (10,993) & 8,147 (16,340) & 7,985 (15,510) & 23,655 (36,571) & 162,751 (181,016) \\ 
  \hline
  0.550 & 0.525 & 0.0 & 0.0 & 1,178 (2,288) & 1,245 (2,512) & 1,533 (3,293) & 1,845 (3,642) & 4,480 (7,347) & 130,302 (158,069) \\ 
  0.550 & 0.525 & 0.0 & 0.5 & 1,116 (2,252) & 1,103 (2,164) & 1,349 (2,970) & 1,637 (3,370) & 4,277 (6,826) & 164,400 (195,030) \\ 
  0.550 & 0.525 & 0.5 & 0.0 & 843 (1,739) & 866 (1,777) & 1,019 (2,183) & 1,120 (2,249) & 4,228 (6,530) & 73,347 (116,466) \\ 
  0.550 & 0.525 & 0.5 & 0.5 & 769 (1,502) & 763 (1,526) & 867 (1,829) & 1,083 (2,140) & 4,109 (6,117) & 66,991 (112,060) \\ 
  \hline
  0.550 & 0.550 & 0.0 & 0.0 & 309 (616) & 313 (643) & 373 (781) & 436 (880) & 1,139 (1,774) & 110,469 (190,030) \\ 
  0.550 & 0.550 & 0.0 & 0.5 & 288 (570) & 294 (585) & 321 (678) & 414 (847) & 1,080 (1,692) & 67,345 (189,989) \\ 
  0.550 & 0.550 & 0.5 & 0.0 & 233 (467) & 235 (483) & 259 (541) & 304 (608) & 1,099 (1,640) & 436 (1,163) \\ 
  0.550 & 0.550 & 0.5 & 0.5 & 206 (417) & 211 (413) & 248 (511) & 279 (563) & 1,078 (1,580) & 415 (978) \\ 
  \hline
  0.600 & 0.550 & 0.0 & 0.0 & 310 (618) & 315 (622) & 354 (755) & 431 (903) & 1,223 (1,902) & 52,712 (98,107) \\ 
  0.600 & 0.550 & 0.0 & 0.5 & 289 (576) & 269 (530) & 310 (629) & 400 (804) & 1,186 (1,790) & 43,011 (89,454) \\ 
  0.600 & 0.550 & 0.5 & 0.0 & 189 (367) & 190 (368) & 210 (408) & 270 (528) & 1,165 (1,698) & 234 (545) \\ 
  0.600 & 0.550 & 0.5 & 0.5 & 179 (355) & 181 (344) & 203 (394) & 232 (471) & 1,136 (1,627) & 233 (510) \\ 
  \hline
  0.600 & 0.600 & 0.0 & 0.0 & 80 (155) & 82 (158) & 90 (171) & 104 (214) & 300 (447) & 130 (298) \\ 
  0.600 & 0.600 & 0.0 & 0.5 & 78 (154) & 79 (158) & 85 (153) & 98 (213) & 300 (443) & 124 (294) \\ 
  0.600 & 0.600 & 0.5 & 0.0 & 59 (114) & 61 (121) & 77 (129) & 82 (156) & 298 (440) & 61 (123) \\ 
  0.600 & 0.600 & 0.5 & 0.5 & 57 (110) & 58 (113) & 73 (129) & 75 (142) & 295 (420) & 60 (122) \\ 
   \hline
\end{tabular}}
\caption{Estimated expected (90th percentile) sample sizes to confirm that the assorter mean is greater than 1/2 at risk limit 5\% for TSMs  with various betting strategies (right columns) for sampling without replacement from ONEAudit overstatement populations with various reported assorter means ($\bar{A}^c$), true assorter means ($\bar{A}^m$) spread between the CVR and batch strata ($\Delta_{\mbox{\footnotesize a}}$), and spread between batches within the batch stratum ($\Delta_{\mbox{\footnotesize w}}$). 
Estimates were made from the sample mean and 0.9 empirical quantile of 1,000 sample sizes simulated by drawing simple random samples without replacement.
Each population consists of 200,000 total cards; 100,000 cards have CVRs and 100,000 cards are in batches, each of size 1,000, modeling precinct-level totals.
Oracle Kelly: optimal bet for sampling with replacement from the actual population of overstatements.
AP Kelly: optimal bet based on sampling with replacement from an assumed (\emph{a priori}) population of overstatements.
Universal Portfolio: discrete approximation to the universal portfolio of Cover \cite{cover91};
see Waudby-Smith and Ramdas \cite{waudby-smithRamdas24} and Waudby-Smith et al.  \cite{waudby2025universal}.
AGRAPA: approximate growth-rate adapted to the particular alternative bet \cite{waudby-smithRamdas24}.
Truncated shrinkage: bet set implicitly by a truncated shrinkage estimate of the population mean \cite{stark23} using the default parameters in SHANGRLA \url{https://github.com/pbstark/SHANGRLA}.
COBRA: comparison-optimal bet \cite{spertus2023cobra} for sampling with replacement, computed on the assumption that there are no one-vote overstatements and 2-vote overstatements occur for 0.1\% of cards.
The columns are sorted in rough order of increasing average sample size.
}
 \label{tab:betting_stopping_times}
\end{table}

\section{Case study: San Francisco 2024 Mayoral race}
\label{sec:case_study}

We consider the 2024 mayoral race in San Francisco as a case study. 
This instant-runoff voting (IRV) contest included thirteen candidates.
Daniel Lurie, who received 26\% of the first-choice selections and 55\% after all but two candidates were eliminated, defeated incumbent London Breed, who received 24\% of the first-choice elections and 45\% of the final round votes.
All computations for this IRV election were carried out using the SHANGRLA codebase, to which we contributed new functions for testing-by-betting. 
The Jupyter Notebook we used to demonstrate the IRV ONEAudit and estimate workloads under different bets is available at \url{https://github.com/spertus/UI-TS/blob/main/Code/SF_oneaudit_example.ipynb}.

The election produced 1,603,908 CVRs, of which 216,286 were for cards cast in 4,223 precinct batches and 1,387,622 CVRs were for vote-by-mail (VBM) cards.
VBM CVRs are linked to the corresponding card, facilitating ballot-level comparison auditing,
but the in-person CVRs are not linked to individual cards, only to tabulation batches.
The CVRs were incorporated into the audit using ONEAudit.
RAIRE \cite{blom18} was used to generate the assertions for the audit to test.
The 80th percentile of the distribution of sample sizes for each of the five bets introduced above was estimated from the 0.8 empirical quantile of sample sizes required to confirm the winner in 100 simulated audits with no error in the reported tallies. 
Results appear in Table~\ref{tab:sf_results}. 
Kelly optimal and AGRAPA reduced the sample size by about half compared to the truncated shrinkage estimate bet or the COBRA bet.

\begin{table}[htbp]
\centering
\scalebox{0.7}{
\begin{tabular}{|c|c|c|c|c|}
  \hline
 Kelly-optimal & AGRAPA & Universal Portfolio & truncated shrinkage & COBRA \\ 
\hline
 111 & 111 & 165 &  192 & 211 \\ 
   \hline
\end{tabular}}
\caption{
Estimated 80th percentile sample sizes of ONEAudit RLAs at a 5\% risk limit using betting TSMs with various betting strategies computed under sampling without replacement from the ONEAudit overstatement population corresponding to the 2024 San Francisco Mayoral race, which used instant-runoff voting (IRV).
Estimates are the 0.8 empirical quantile of 100 sample sizes simulated by drawing simple random samples of cards without replacement and treating the CVR for each card as if it were an accurate manual-vote record (MVR).
For cards cast in precincts, the assorter applied to the MVR is compared to the ONEAudit average assorter value for the batch.
Kelly-optimal: optimal for sampling with replacement from the overstatements implied by the batch-level results, on the assumption that the results are accurate.
AGRAPA: approximate growth-rate adapted to the particular alternative \cite{waudby-smithRamdas24}.
Universal Portfolio: discrete approximation to Cover \cite{cover91};
see Waudby-Smith and Ramdas \cite{waudby-smithRamdas24} and Waudby-Smith et al. \cite{waudby2025universal}.
Truncated shrinkage: bet set implicitly by a truncated shrinkage estimate of the population mean \cite{stark23} using the default parameters in SHANGRLA \url{https://github.com/pbstark/SHANGRLA}.
COBRA: comparison-optimal bet \cite{spertus2023cobra} on the assumption that there are no one-vote overstatements and a 0.1\% rate of 2-vote overstatements.
}
 \label{tab:sf_results}
\end{table}

\section{Discussion}
\label{sec:discussion}

ONEAudit is simpler and more efficient without stratification.
ONEAudit can be made substantially more efficient by employing approximately Kelly-optimal or universal portfolio strategies for betting.
The universal portfolio can be quickly and sufficiently approximated by a discrete mixture of betting TSMs over a grid of bets, which has already proven useful for card-polling and card-level comparison audits.
Simulations and data from a 2024 IRV contest in San Francisco, CA, strongly suggest that unstratified ONEAudit with \emph{a priori} Kelly optimal betting or discrete portfolio betting are simple and efficient methods to conduct RLAs of elections that use diverse voting technology.

There remain gaps to be explored.
We did not determine how much tally error would be necessary to make \textit{a priori} Kelly betting worse than the universal portfolio or some other strategy that does not rely on the reported results.
We also only simulated elections with tally errors in batches, not in CVRs. 
We expect errors in the batches to have less effect on efficiency, but also to be more common in real elections. 
In practice, it could be desirable to ensure both efficiency when the reported tallies are accurate and robustness to possible tally errors by mixing an \textit{a priori} Kelly TSM and a universal portfolio TSM.
The mixture could start with high weight (e.g., 0.9) on the \textit{a priori} Kelly TSM and, as the sample size increases, shift weight to the universal portfolio, which asymptotically attains nearly the same growth rate as the oracle Kelly TSM.
Alternatively, the universal portfolio could use a weight function in the quadrature that puts more weight on the value of $\lambda$ corresponding to the \emph{a priori} Kelly bet instead of uniform weights.
These ideas could easily be implemented in the SHANGRLA codebase. 

We only examined bets optimized for sampling with replacement, and therefore did not update as the conditional null mean changed. 
This makes it easier to compute the Kelly TSM---it only needs to be computed once---but the bets could be adjusted to account for sampling without replacement. 
We expect the effect of this change to be minor: the marginal improvement only becomes noticeable once the sample includes a substantial portion of the population, at which point a full hand count is generally more efficient than random sampling.

It is possible that stratification would increase efficiency if the strata partitioned batches by error rate, a case we did not examine.
We speculate that it would take quite high error rates and that the strata would need to discriminate accurately between low and high error rates for stratification to make much difference. 
If auditors believed they had such information, a different kind of inspection (e.g., a forensic audit) might be more appropriate, since high error rates should occur only if there were gross procedural errors, misconfiguration, or malfunction.
As we see it, with ONEAudit and universal-portfolio betting (possibly augmented by shrinking towards the \emph{a priori} Kelly bet), the only reason for stratification is to accommodate legal requirements---not to reduce the overall workload or to accommodate diverse election equipment.

\bibliographystyle{splncs04} 
\bibliography{RLA_bib.bib}

\end{document}